\documentclass[%
reprint,
showpacs,
amsmath, amssymb,aps,
]{revtex4-1}

\usepackage{graphicx}
\usepackage{dcolumn}
\usepackage{bm}
\usepackage{color}

\newcommand{\md}{\mathrm d}

\begin{document}


\title{A toy model for weak interaction
based on condensed gauge bosons}%

\author{Hiroaki Kohyama}
\affiliation{Department of Physics,
National Taiwan University, Taipei 10617, Taiwan}

\date{\today}

\begin{abstract}
We construct a toy model for weak interaction based on the
assumption that gauge bosons form condensates. We then
discuss the model predictions calculated from the effective
Feynman rules which are derived through computing the
effective action.
\end{abstract}


\maketitle

\section{\label{sec:intro}%
Introduction}
It is known that physics relating to the electroweak interactions
is well described by the standard model
(SM)~\cite{Weinberg:1967tq, Salam:1968rm}
in which Higgs particle plays an important role.

In this paper, from mathematical interest, we construct a toy
model based on the assumption of condensed gauge bosons.
The motivation to consider the gauge boson condensation 
is on the fact that such hypothetical condensate can give the
effective masses to the gauge bosons.
Here we first construct a toy model then discuss the properties
of the obtained model. We believe that this type of speculations
is useful for considering possible model constructions and their
restrictions by experiments.

\section{\label{sec:TM}%
Toy model}
The Lagrangian for the ${\rm SU}(2)$ gauge field coupled to
fermions is given by
\begin{align}
  \mathcal{L}
  =
  & \overline{\psi} (i\partial\!\!\!/ -m) \psi
   -\frac{1}{4} (\partial_\mu A^a_\nu - \partial_\nu A^a_\mu)^2
     \nonumber \\
  & +g \overline{\psi} \gamma^\mu \tau^a \psi A^a_\mu 
     -g f^{abc}(\partial_\mu A_\nu^a) A^{\mu b} A^{\nu c}
     \nonumber \\
  & -\frac{1}{4} g^2
        (f^{eab} A^{a}_\mu A^{b}_\nu)
        (f^{ecd} A^{\mu c} A^{\nu d}),
\end{align}
where $\psi$ is the fermion doublet field, $\psi= (\nu_e, e)^T$,
$m$ is the diagonal mass matrix, $m={\rm diag}(m_{\nu_e},m_e)$,
$A_\mu^a$ is the ${\rm SU}(2)$ gauge field, $g$ is the coupling constant,
$\tau^a=\sigma^a/2$ with the Pauli matrices $\sigma^a$ and
$f^{abc} = i\epsilon^{abc}$
is the structure constant for ${\rm SU}(2)$ gauge sector being the
antisymmetric tensor.

By using the following notations,
\begin{align}
  W^\pm_\mu = \frac{1}{\sqrt{2}}(A^1_\mu \mp i A^2_\mu),
  \quad
  Z_\mu = A^3_\mu,   
\end{align}
the Lagrangian can be rewritten as
\begin{equation}
  \mathcal{L}
  =
   \overline{\psi} (i\partial\!\!\!/ -m) \psi
     +\mathcal{L}_{\rm K}
     +\mathcal{L}_{\rm I}
     +\mathcal{L}_{3}
     +\mathcal{L}_{4},
\end{equation}
with the kinematic term
\begin{align}
  \mathcal{L}_{\rm K}
  &=  - (\partial_\mu W^+_\nu )( \partial^\mu W^{-\nu})
         + (\partial_\mu W^+_\nu )( \partial^\nu W^{-\mu})
     \nonumber \\
  & \quad
   -\frac{1}{4} (\partial_\mu Z_\nu - \partial_\nu Z_\mu)^2,
\end{align}
the interaction with fermions,
\begin{align}
  \mathcal{L}_{\rm I}
  &=\frac{g}{\sqrt{2}} \overline{\nu}_e \gamma^\mu e W^+_\mu
      +\frac{g}{\sqrt{2}}  \overline{e} \gamma^\mu \nu_e W^-_\mu 
      \nonumber \\
  & \quad +\frac{g}{2} \overline{\nu}_e \gamma^\mu \nu_e Z_\mu
      - \frac{g}{2} \overline{e} \gamma^\mu e Z_\mu ,
\end{align}     
and the interacting terms for gauge bosons,
\begin{align}     
 \mathcal{L}_{\rm 3}
  &= -g \bigl[ (\partial_\mu W_\nu^+)
            ( W^{- \mu} Z^{\nu} - Z^\mu W^{- \nu})
     \nonumber \\
  & \quad + (\partial_\mu W_\nu^-)
            ( Z^\mu W^{+ \nu}  - W^{+ \mu} Z^\nu)
     \nonumber \\
  & \quad + (\partial_\mu Z_\nu)
            ( W^{+\mu} W^{- \nu}  - W^{- \mu} W^{+\nu})
            \bigr],  \\
 \mathcal{L}_{\rm 4}
  &= \frac{1}{2} g^2
        ( W^{+}_\mu W^{-\mu} )^2
        -\frac{1}{2} g^2
        ( W^{+}_\mu W^{+\mu} )
        ( W^{-}_{\nu} W^{-\nu})
        \nonumber \\
  &\quad 
        +g^2
         ( W^{+}_\mu W^{-\mu} ) Z_\nu Z^\nu
        -g^2
        ( W^{+}_\mu Z^{\mu} )
        ( W^{-}_{\nu} Z^{\nu}).
\end{align}

What is actually observed in experiments are the $n$-point
functions which should be calculated from the effective
Feynman rules derived from the effective action, $\Gamma =
\int \md^4x \langle \mathcal{L} \rangle + \cdots$, e.g., 
\begin{align}
  iD_{Z}^{-1}(x,y)
   &= \frac{\delta^2 \Gamma}
                 {\delta Z_\mu(x) \delta Z_\nu(y)},
\end{align}
then one has to evaluate the effective action for the sake of
deriving the model predictions~\cite{Peskin:1995ev}. In evaluating
the effective action, we assume that the gauge boson fields have
non-zero expectation values,
\begin{align}
  \langle W^+_\mu W^-_\nu \rangle = g_{\mu \nu} \phi_W,
  \quad
  \langle Z_\mu Z_\nu \rangle = g_{\mu \nu} \phi_Z,
\end{align}
where we call $\phi_W$ and $\phi_Z$ as the gauge boson
condensates. The motivation of speculating this hypothesis
comes from the Bose Einstein condensate at extremely low
temperature in condensed matter physics \cite{Kohyama:2016dhd}.
Due to the non-zero values of the condensates, there appear
the mass terms in the effective propagators for the gauge
bosons, which will be discussed in the next section.

\section{\label{sec:EM}%
Effective masses for gauge bosons}
Performing the functional derivative of the effective action,
one obtains the following effective propagators in the Feynman
gauge,
\begin{align}
  D_{W}(p) = \frac{-ig^{\mu \nu}}
                 {p^2 - M_W^2},
  \quad
  D_{Z}(p) = \frac{-ig^{\mu \nu}}
                 {p^2 - M_Z^2},
\end{align}
where
\begin{align}
  M_W^2 = 3g^2 (\phi_W + \phi_Z), 
  \quad
  M_Z^2 = 6g^2 \phi_W.
\label{mass}
\end{align}
From the experiments we have 
$M_W = 80.2{\rm GeV}$,
$M_Z = 91.2{\rm GeV}$,
then we see
$g^2 \phi_W = (37.2{\rm GeV})^2$,
$g^2 \phi_Z = (27.5{\rm GeV})^2$.
Thus the expectation values of $\phi_W$ and $\phi_Z$ are
different; here we think the difference stems from the
different electric charge between $W$ and $Z$ bosons.

\section{\label{sec:interaction}%
Fermion interactions}
It is known that the left and right handed fermions interact
differently with the gauge bosons. This means that the effective
couplings, $g^{\rm L}_{W}$ and $g^{\rm R}_{W}$ for left and
right handed interactions calculated from 
\begin{align}
  \frac{\delta^3 \Gamma}
          {\delta  \overline{\nu}_{e {\rm L(R)}}  \delta e_{\rm L(R)} 
           \delta W_\mu^+}
  \to \gamma^\mu g_{W}^{\rm L(R)},
\end{align}
lead different values, namely, $g^{\rm L}_{W} \neq g^{\rm R}_{W}$.
Based on the experimental results we usually set $g_{W}^{\rm R}=0$,
which indicates that the $W$ bosons do not effectively interact
with right handed fermions via weak force.

Similarly, the effective couplings for $Z$ boson and $\nu_e$, $e$
can be denoted by $g_{Z\nu_e}^{\rm L}$, $g_{Z\nu_e}^{\rm R}$,
$g_{Ze}^{\rm L}$ and $g_{Ze}^{\rm R}$, where we also set
$g_{Z \nu_e}^{\rm R} = 0$ for the empirical reason.
Theoretically, it might be possible to evaluate the forms of the
effective couplings through carefully calculating the effective
action. While it is more practical to fix the values of
effective couplings by the experimental observations.

Here, it may be interesting to make the comparison with the
forms appearing in the SM. The counterparts become,
\begin{align}
  &g_{W}^{\rm L} \to \frac{g_{\rm SM}}{\sqrt{2}},
  \,\,
  g_{W}^{\rm R} \to 0,
  \quad
  g_{Z\nu_e}^{\rm L} \to \frac{g_{\rm SM}}{2 \cos \theta_W},
  \,\,
  g_{Z\nu_e}^{\rm R} \to 0,
  \nonumber \\ 
  &g_{Z e}^{\rm L} \to \frac{g_{\rm SM}}{\cos \theta_W}
  \left( -\frac{1}{2} + \sin^2 \theta_W  \right), \,\,
  g_{Z e}^{\rm R} \to \frac{g_{\rm SM}}{\cos \theta_W} \sin^2 \theta_W, 
  \nonumber
\end{align}
where $g_{\rm SM}$ represents the coupling strength in the SM and
$\theta_W$ does the Wineberg angle.

\section{\label{sec:level}%
Possible energy levels for condensates}
Finally, we shall present the further possibility of extending the current
toy model.

We think that the objects formed by the condensates have finite
size in the coordinate space, namely $\phi(t, r) \neq 0$ for finite $r$,
but $\phi(t, r) \to 0$ for $r \to \infty$. As seen in solving the
Schr\"{o}dinger equation in quantum mechanics,
the above restriction leads discreet energy levels of the condensates.
Therefore, we expect that the condensates can have various energy
levels,
\begin{align}
  \phi_W^{(n)}, \quad \phi_Z^{(n)}, 
\end{align}
with energies which would give different masses to gauge bosons where
the straightforward generalization of Eq.~(\ref{mass}) become
\begin{align}
  {M_W^{(n,m)}}^2 = 3g^2 (\phi_W^{(n)} + \phi_Z^{(m)}), 
  \quad
  {M_Z^{(n)}}^2 = 6g^2 \phi_W^{(n)}.
\end{align}
Thus, we believe that the extended version of our model has the
possibility of describing the properties of the gauge bosons with
heavier masses.

The energy levels of the condensates can be, in principle, evaluated
by solving the Schr\"{o}dinger equation if one can calculate the potential
derived from the non-Abelian gauge interaction.

\section{\label{sec:conclusion}%
Concluding remarks}
We have considered the toy model by assuming the condensed
gauge bosons. We find that the gauge bosons can have effective
masses if the condensation occurs. We believe that the
speculation presented here is also useful in considering
models for QCD.

\begin{acknowledgments}
The author is supported by Ministry of Science and Technology
(Taiwan, ROC), through Grant No. MOST 103-2811-M-002-087.
\end{acknowledgments}




\end{document}